\documentclass[aps, prl, twocolumn,showpacs,showkeys,preprintnumbers,amsmath,amssymb,superscriptaddress]{revtex4}
\usepackage{graphicx}
\usepackage{natbib}
\usepackage{color}
\usepackage{amsmath}
\usepackage{enumerate}
\usepackage[colorlinks=true,linkcolor=blue,urlcolor=blue,citecolor=blue]{hyperref}
\usepackage[toc,page]{appendix}
\usepackage[normalem]{ulem}
\def\strutdepth{\dp\strutbox}
\def\nw#1{\strut\vadjust{\kern-\strutdepth\vtop to0pt{\vss\hbox to\hsize
{\hskip\hsize\hskip5pt$\leftarrow$\hss\strut}}}{\em #1}}

\newcommand{\tEff}{\tau_{\mbox{\scriptsize{eff}}}}

\newcommand{\peq}{p_{eq}}

\begin{document}

\title{Kovacs Effect in Glass with Material Memory Revealed in Non-Equilibrium Particle Interactions}

\author{Matteo Lulli}
\affiliation{Department of Mechanics and Aerospace Engineering, Southern University of Science and Technology, Shenzhen, Guangdong 518055, China}
\author{Chun-Shing Lee}
\affiliation{Department of Applied Physics, Hong Kong Polytechnic University, Hong Kong, China}
\author{Ling-Han Zhang}
\affiliation{Department of Physics, Carnegie Mellon University, Pittsburgh, Pennsylvania 15213, USA}
\author{Hai-Yao Deng}
\affiliation{School of Physics and Astronomy, Cardiff University, 5 The Parade, Cardiff CF24 3AA, United Kingdom}
\author{Chi-Hang Lam}
\email{C.H.Lam@polyu.edu.hk}
\affiliation{Department of Applied Physics, Hong Kong Polytechnic University, Hong Kong, China}

\date{\today}

\begin{abstract}
    The Kovacs effect is a remarkable feature of the ageing dynamics of glass forming liquids near the glass transition temperature. It consists in a non-monotonous evolution of the volume/enthalpy after a succession of two abrupt temperature changes: first from a high initial temperature $T_i$ to a much lower annealing temperature $T_a$ followed by a smaller second jump back to a slightly higher final temperature $T_f$. The second change is performed when the instantaneous value of the volume/enthalpy coincides with the equilibrium one at the final temperature. While this protocol might be expected to yield equilibrium dynamics right after the second temperature change, one observes the so-called Kovacs hump in glassy systems. In this paper we apply such thermal protocol to the Distinguishable Particles Lattice Model (DPLM) for a wide range of fragility of the system. We study the Kovacs hump based on energy relaxation and all main experimental features are captured. Results are compared to general predictions based on a master equation approach in the linear response limit. We trace the origin of the Kovacs hump to the non-equilibrium nature of the probability distribution of particle interaction energies after the annealing and find that its difference with respect to the final equilibrium distribution is non-vanishing with two isolated zeros. This allows Kovacs' condition of equilibrium total energy to be met out-of-equilibrium, thus representing the memory content of the system. Furthermore, the hump is taller and occurs at a larger overlap with the system initial configuration for more fragile systems. The dynamics of a structural temperature for the mobile regions strongly depends on the glass fragility while for the immobile ones only a weak dependence is found.
\end{abstract}

\maketitle

\section{Introduction}\label{sec:introduction}
\begin{figure}[!ht]
  \includegraphics[scale=0.42]{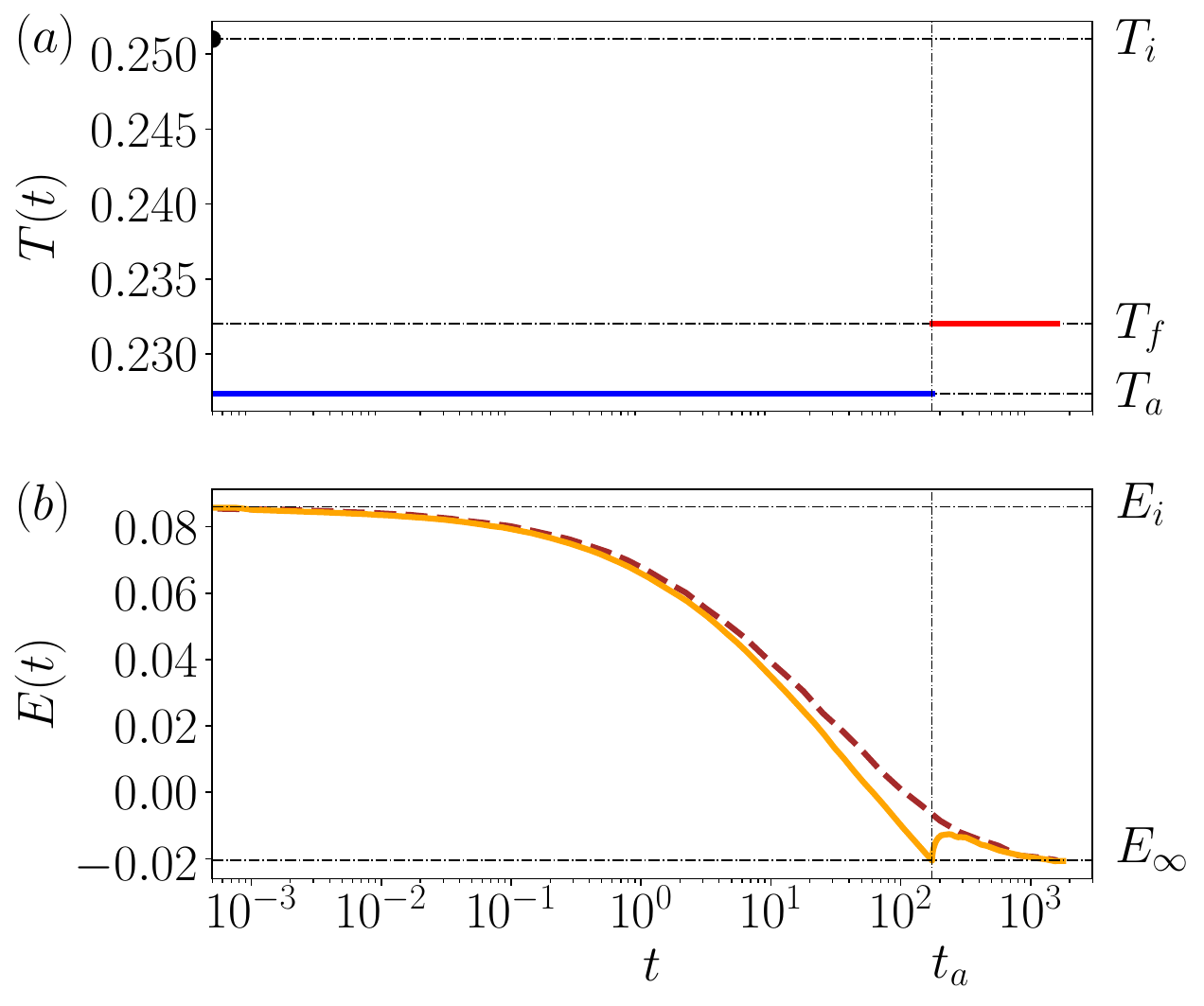}
  \caption{Schematic presentation of the memory protocol for $G_0 = 0.1$: Panel (a) and (b) report the three stages of the protocol for the temperature and energy respectively. Stage 1: The system starts from equilibrium at the initial temperature $T_i$ (see values in Table~\ref{tab:fragilities}). Stage 2: At time $0$ the first temperature jump to $T_a < T_i$ (down-jump) is performed and the energy relaxes. Stage 3: At time $t_a$ the instantaneous value of the energy coincides with the equilibrium one at the final temperature $T_f$, i.e. $E(t_a) = E_\infty$ and the second temperature jump to $T_f > T_a$ (up-jump) is performed. The energy, represented by the solid line, has a nonmonotonic behavior passing through a maximum and relaxing back to the final equilibrium value. The dashed line reports the energy relaxation for a single down-jump directly to $T_f$.}\label{fig:memory_sketch}
\end{figure}

Nonlinearities in the aging dynamics are a hallmark of glassy systems. Kovacs' series of experiments~\cite{Kovacs1964} is one of the cornerstones on which our present understanding of glassy dynamics is rooted~\cite{Hodge1994, Angell2000Rev, Roth2016}. Kovacs' work \cite{Kovacs1964} thoroughly analyzed the volume relaxation dynamics of polymer glasses (PVAc) by performing abrupt temperature changes, or \emph{temperature jumps}. Two important results of these analyses are the renowned \emph{Kovacs effect}~\cite{Berthier2002_sg, Bertin2003_Kovacs, Cugliandolo2004, Mossa2004, Buhot2003, Arenzon2004, Aquino2008, Prados2010_KovacsME} and the \emph{expansion gap paradox}~\cite{McKenna1995, McKenna1999, Kolla2005, Hecksher2010, Hecksher2015, Banik2018, Struik1997_1, Struik1997_2}, for double- and single-temperature jumps respectively. The expansion gap paraodox refers to an apparent difference in the instantaneous relaxation time $\tEff$ near equilibrium, between heating (up-jump) and cooling (down-jump), after a single temperature change is performed: while for the down-jump case the values of $\tEff$ converge at equilibrium, independently on the initial temperature, for the up-jump case the data display a dependence on the initial condition even near the end of the relaxation. On the other hand, the \emph{Kovacs effect} shows how the instantaneous value of the volume (or enthalpy~\cite{Montserrat1994,Grassia2018}) during the relaxation, is not a sufficient indicator of the departure from equilibrium of the system. After a first temperature jump from the initial temperature $T_i$ to the annealing temperature $T_a$ (with $T_a < T_i$), the relaxation is interrupted when the observable reaches the equilibrium value of a third final temperature $T_f$, identifying the annealing time $t_a$. A second temperature jump from $T_a$ to $T_f$ (with $T_a < T_f < T_i$) is then performed. One observes a non-monotonous evolution, rather than a constant one, of the already-attained equilibrium value (see Fig.~\ref{fig:memory_sketch}). The \emph{Kovacs effect} has been studied by means of finite-dimensional and mean-field spin-glass models \cite{Berthier2002_sg, Cugliandolo2004, Krzakala2006}, ordered XY and Ising models \cite{Berthier2002_xy, Bertin2003_Kovacs, Krzakala2006}, molecular dynamics \cite{Mossa2004}, kinetically constrained models (KCMs)~\cite{Buhot2003, Arenzon2004} and simple two-level systems \cite{Aquino2008, Prados2010_KovacsME}. Also mean-field constitutive models have been used, most notably the Tool-Narayanaswamy-Moynihan (TNM)~\cite{Tool1946, Narayanaswamy1971, Moynihan1976} and the Kovacs-Aklonis-Hutchinson-Ramos (KAHR) models~\cite{KAHR1979}, and those accounting for fluctuations of observables such as a stochastic version of a free-volume model~\cite{Robertson1984} and the Stochastic Constitutive Model (SCM)~\cite{Medvedev2015}. Recently, the Kovacs effect has also been investigated in granular fluids~\cite{Prados_2014, Trizac_2014}, disordered mechanical systems~\cite{Lahini_2017} and in active matter suspensions~\cite{K_rsten_2017} demonstrating how such memory effect offers an interesting window into the dynamics of a wide variety of physical systems.

We reproduce all the features of Kovacs effect using the recently developed Distinguishable Particles Lattice Model (DPLM)~\cite{DPLM2017}. The phonon temperature, which is subjected to two consecutive jumps, is modeled by the bath temperature of the kinetic Monte Carlo simulation of the DPLM. We observe the characteristic \emph{Kovacs hump} during the system energy relaxation, analogous to enthalpy relaxation in experiments~\cite{Montserrat1994, Bernazzani_2002, Grassia2018}. We study several features and rescaling properties of the Kovacs hump~\cite{Cugliandolo2004, Prados2010_KovacsME, Prados_2014} while probing the linear response regime. We are able to identify the memory content of the system by means of the instantaneous distribution of the particle interaction energies $p(V,t)$. In particular we identify the out-of-equilibrium features of $p(V, t)$ at time $t_a$ after the annealing by means of the difference with respect to the final equilibrium distribution, i.e. $p(V, t_a) - \peq(V, T_f)$. The latter quantity displays two zeros allowing the system to reach the same final equilibrium value of the energy while being out-of-equilibrium. Furthermore, a recent version of DPLM~\cite{Lee_2020} allows us to investigate the interplay between Kovacs effect and fragility~\cite{Aquino_2006}: as the fragility increases, the hump height increases and so is the associated fraction of particles retaining the initial positions, i.e., the overlap $q$~\cite{Lulli_2020}. Finally, by parametrizing the system evolution by means of $q$, we can clearly show fragility-dependent features of the dynamics based on a structural temperature~\cite{Lulli_2020}.

Recent works show that the DPLM displays many features of the particle dynamics of glass formers and offers the possibility of performing exact equilibrium calculations of the free-energy~\cite{DPLM2017, Lee_2020}. In particular, the DPLM is the first particle model to successfully reproduce, the \emph{expansion gap paradox}~\cite{Lulli_2020}, providing an intuitive explanation in terms of a local structural temperature whose dynamics is spatio-temporally unstable in the up-jump case. Furthermore, the DPLM allows for controlling the \emph{kinetic fragility}~\cite{Lee_2020} over a very wide range of values of the fragility index. The relation between kinetic and thermodynamic fragility correctly captures several experimental features as well. It has also been possible to obtain an analytical expression for the particles Mean Squared Displacement (MSD)~\cite{lam2018, Deng2019} which is in good agreement with simulation results.

The results of this paper convey a comprehensive picture of the ability of the DPLM to reproduce experimental signatures of glassy systems observed in Kovacs' experiments~\cite{Kovacs1964}. It is a major challenge to study a wide range of glassy phenomena in a unified framework based on a consistent set of assumptions. We demonstrate that, while correctly reproducing the expansion gap~\cite{Lulli_2020}, the DPLM is also able to capture the Kovacs effect. To the best of our knowledge only phenomenological constitutive models, namely the SCM~\cite{Medvedev2015} and the free-volume model~\cite{Robertson1984}, have been shown to reproduce both effects.

\section{Model Definition}\label{sec:model}
We simulate the DPLM~\cite{DPLM2017, Lee_2020} defined on a two-dimensional lattice of linear size $L=100$ where the sites are occupied by $N$ distinguishable particles, each of them associated to a \emph{unique label} $s_i = 1,\ldots,N$. An important property of the model is that each particle is coupled to its nearest neighbors by means of \emph{label-dependent} random interactions: considering the interaction energy associated to the particles sitting at sites $i$ and $j$, with labels $s_i$ and $s_j$, one has a two-indices quantity $V_{s_i s_j}$. In order to simulate the hopping of particles we consider the presence of $N_v = 100$ voids, i.e. given that $L^2=N+N_v$, one has a void density $\phi_v = N_v/L^2 = 0.01$. One can write the energy of the system as
\begin{equation}
  \label{eq:E}
  E = \sum_{\langle ij \rangle'} V_{s_i s_j},
\end{equation}
where the sum $\sum_{\langle ij \rangle'}$ is restricted to the couples of neighboring sites occupied by particles only. The entire set of all possible couplings $\{V_{kl}\}$ is drawn according to an \emph{a priori} probability distribution $g(V)$ and it is \emph{quenched}, whereas the set of the \emph{realized} interactions $\{V_{s_i s_j}\}$, i.e. the interactions energies in a given configuration of particles on the lattice, is distributed according to a different function which at equilibrium is proved  and numerically verified to be~\cite{DPLM2017, Lee_2020, Lulli_2020}
\begin{equation}
  \label{eq:peq}
  \peq(V,T) = \frac{1}{\mathcal{N}(T)}\,g(V)\,e^{-V/k_BT},
\end{equation}
where $\mathcal{N}(T)$ is a normalization constant. Several choices are possible for the \emph{a priori} distribution $g(V)$ and in this work we adopted the same model as in \cite{Lee_2020}, and hence $g$ is given by
\begin{equation}\label{eq:g}
    g(V) = \frac{G_0}{V_1 - V_0} + (1 - G_0)\delta(V - V_1)\, ,\, V \in [V_0, V_1],
\end{equation}
with $V_1=-V_0=0.5$.
Here, $G_0 \in (0, 1]$ is a thermodynamic parameter for tuning the fragility \cite{Lee_2020}. For $G_0 = 1$, $g(V)$ reduces to a simple uniform distribution which makes the system a strong glass, while when $G_0$ tends to zero, the system becomes increasingly fragile \cite{Lee_2020}. In the following we adopt natural units, hence $k_B = 1$. The equilibrium sampling is performed by means of a kinetic Monte Carlo approach using Metropolis dynamics. Each particle can hop to the position of a neighboring void, more precisely representing a recently identified quasi-particle referred to as a quasivoid~\cite{Yip_2020}, with a rate
\begin{equation}
  \label{eq:w}
 w = \begin{cases}
        w_0 \exp\left(-\Delta E / T\right) & \text{for $\Delta E > 0$,} \\
        w_0 & \text{for $\Delta E \leq 0$,}
    \end{cases}
\end{equation}
where $\Delta E$ is the energy change of the system induced by the hop. We set $w_0 = 10^6$. 

\begin{table}[t!]
  \centering
  \begin{ruledtabular}
  \renewcommand{\arraystretch}{1.6}
  \begin{tabular}{ccccc}
    $G_0$ & $1.0$ & $0.7$ & $0.3$ & $0.1$ \\
    \hline
    $m_k$ & 7.52 & 9.07 & 13.05 & 18.97 \\
    $m_t$ & 0.93 & 0.99 & 1.49 & 2.26 \\
    \hline
    $T_g$ & 0.149 & 0.199 & 0.246 & 0.228 \\
    \hline
    $T_i$ & 0.165 & 0.219 & 0.270 & 0.251 \\
    $T_f$ & 0.152 & 0.202 & 0.249 & 0.232 \\
    \hline
    \hline
    $\beta$ & 0.605(3) & 0.598(2) & 0.550(1) & 0.461(2) \\
    $\tau$ & 13.3(1) & 14.5(1) & 20.7(1) & 34.8(3) \\
  \end{tabular}
  \end{ruledtabular}
  \caption{In the first three rows we report the values of the kinetic and thermodynamic fragility indices $m_k$ and $m_t$ as well as $T_g$ as obtained in~\cite{Lee_2020}. The initial and final temperatures are denoted by $T_i$ and $T_f$. The stretching exponent $\beta$ and the characteristic time $\tau$ in the last two rows are the results of the fits of the normalized single down-jump relaxation $\varphi(t)$ (see Fig.~\ref{fig:dj}) to the KWW function~\eqref{eq:kww}.}\label{tab:fragilities}
\end{table}

\section{Simulations Results and Analysis}\label{sec:results}
In this work we study the relaxation dynamics of the system energy~\eqref{eq:E} which is akin to the enthalpy relaxation~\cite{Montserrat1994,Grassia2018}. Averages have been computed over a few thousands independent runs with different initial random seeds. The two-temperatures protocol is somewhat more convoluted than the single-jump one, and it is reported for clarity in Fig.~\ref{fig:memory_sketch}. The protocol can be divided into three stages: during the first stage, for $t<0$, the system is at equilibrium at the initial temperature $T_i$, with a constant energy $E_i$; the second stage begins at $t = 0$ when the bath temperature is changed to the annealing value $T_a$, with $T_a < T_i$ (down-jump), and the energy decreases gradually; the third stage begins at $t_a$ when the energy reaches the value $E(t_a) = E_\infty$, in coincidence with the equilibrium value at the final temperature $T_f$, and the temperature is changed once more to the final value $T_f$, with $T_a < T_f$ (up-jump) and $T_f < T_i$. The hallmark of glassy dynamics in the third stage consists in the non-monotonous evolution of the energy that reaches a maximum, i.e. the \emph{Kovacs hump}, and relaxes back to the same value already attained at $t_a$, rather than remaining as a constant at the equilibrium value $E_\infty$ for the final temperature $T_f$.
\begin{figure}[!t]
  \includegraphics[scale=0.43]{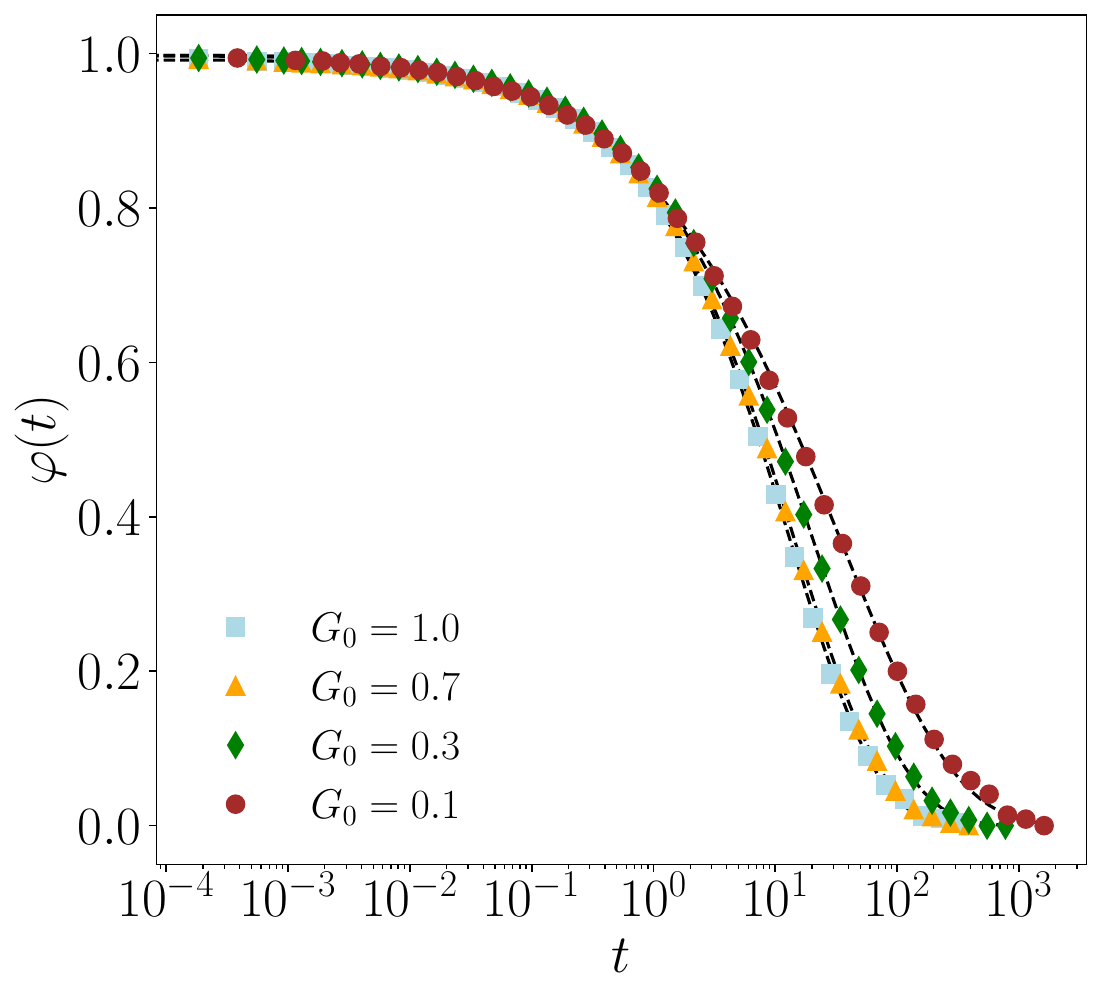}
  \caption{Normalized energy relaxation during down-jumps with fitting to the KWW function~\eqref{eq:kww} (dashed lines). Fitted parameters are reported in Table~\ref{tab:fragilities}.}\label{fig:dj}
\end{figure}
\begin{figure*}[!t]
  \includegraphics[scale=0.24]{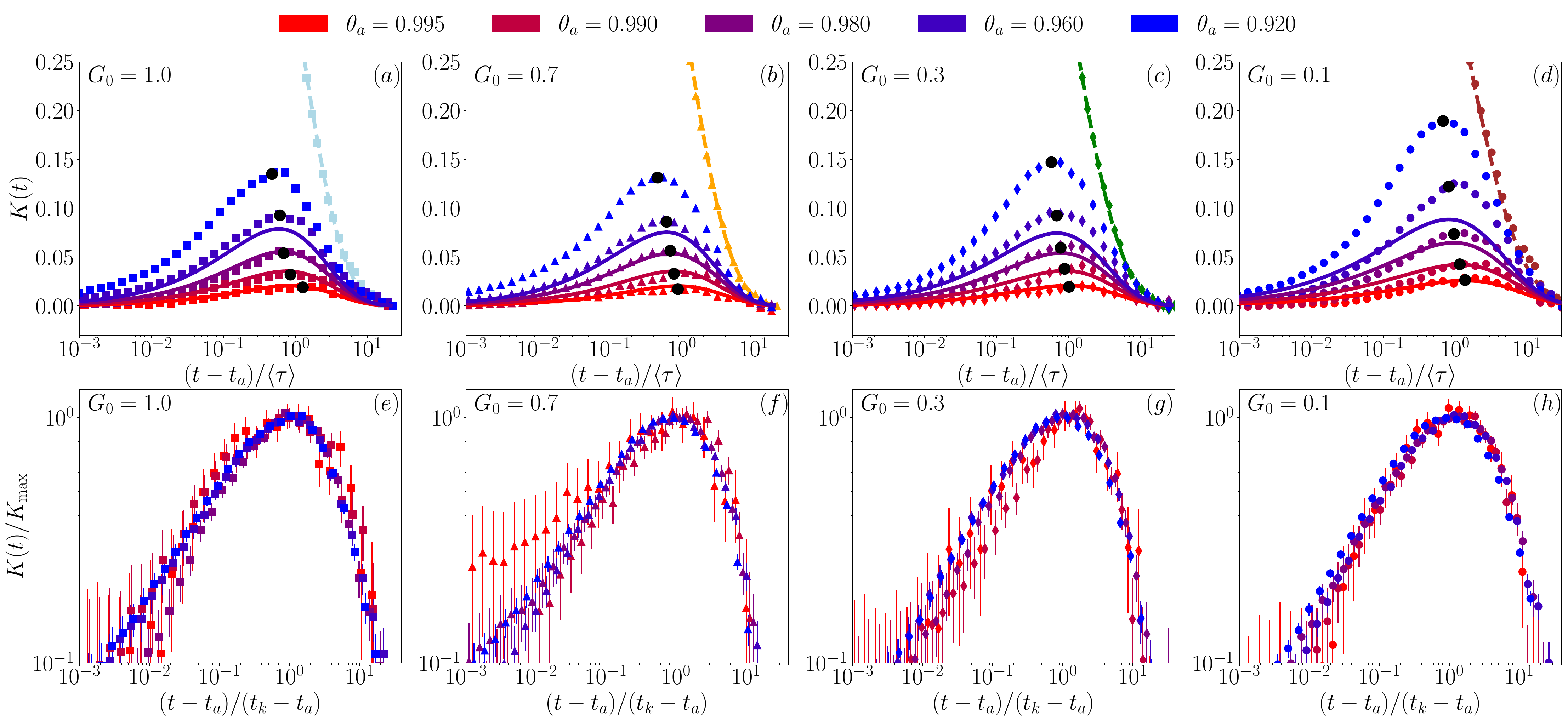}
  \caption{Panels $(a)$ to $(d)$: Kovacs hump for four different values of $G_0$, while those connected by dashed lines represent the single down-jump data. The highest peak corresponds to the lowest annealing temperature $T_a = \theta_a T_f$. The linear response prediction is reported as solid lines (only for $\theta_a \geq 0.960$). Time is rescaled by $\langle \tau \rangle=4 \tau / \pi$ which depends only on $G_0$. Panels $(e)$ to $(h)$: Kovacs hump normalized by its maximum $K_{\text{max}}$ versus normalized time $(t - t_a)/(t_k - t_a)$ with $t_k$ being the time of the peak: data for different annealing temperature mostly superpose onto a single curve, with some deviations due to a strong sensitivity on even a small discrepancy from Kovacs' condition of $E(t_a) = E_\infty$ for the highest $T_a$.}\label{fig:humps}
\end{figure*}

We set the initial-to-glass temperature ratio $T_i/T_g \simeq 1.10$ and the final-to-glass temperature ratio $T_f / T_g \simeq 1.01$, with $T_g$ the fragility-dependent glass transition temperature~\cite{Lee_2020}. The ratios are chosen to match the values used in experiments~\cite{Bernazzani_2002}. Finally, for the annealing temperature $T_a$ we select five different anneal-to-final temperature ratios $\theta_a = T_a/T_f \in \{0.995, 0.990, 0.980, 0.960, 0.920\}$ in order to explore the linear response regime of the system. Given the final-to-glass temperature ratio $T_f / T_g \simeq 1.01$ it follows that the last three values of the annealing temperature are below $T_g$. As for the tuning of the fragility we use four different values of the parameter $G_0 \in \{1.0, 0.7, 0.3, 0.1\}$ yielding a fairly wide range of variation. We report in Table~\ref{tab:fragilities} the corresponding values for the kinetic ($m_k$) and thermodynamic ($m_t$) fragility indices as well as the glass transition temperature $T_g$. As detailed in~\cite{Lee_2020}, $T_g$ is computed as the temperature at which the particles diffusion coefficient matches the reference value $D_r = 10^{-1}$, the smallest value we can adopt in our simulations. However, smaller and more realistic values of $D_r$ would yield a larger fragility, compatible with the experimental scale. The simulations of the memory protocol are set up by first performing single down-jump simulations from $T_i$ to $T_a$ in order to interpolate the time value $t_a$ at which the average energy matches the equilibrium value at $T_f$, i.e. $E(t_a) = E_\infty$. Then, the simulations are restarted keeping the same initial random seeds and using the two-temperatures protocol so that at $t_a$ we set $T = T_f$ in order to analyze Kovacs' hump. This procedure allows us to analyze the linear response regime~\cite{Prados2010_KovacsME, Ruiz_Garc_a_2014} which is very sensitive to discrepancies between $E(t_a)$ and $E_\infty$.

\subsection{Single-jump relaxation}
Let us begin by studying the properties of the single down-jump relaxation for the four different values of $G_0$. Following~\cite{Prados2010_KovacsME, Ruiz_Garc_a_2014} we use the fractional departure from equilibrium defined as the ratio
\begin{equation}\label{eq:phi}
  \varphi(t) = \frac{E(t) - E_\infty}{E(0) - E_\infty},
\end{equation}
where $E_\infty$ is the equilibrium energy at the final temperature $T_f$ and $E(t)$ is the instantaneous value according to~\eqref{eq:E}. Given its definition it follows that $0\leq \varphi(t)\leq 1$. We fit $\varphi(t)$ by means of the Kohlrausch-Williams-Watts (KWW) function~\cite{Kohlrausch_1854, Kohlrausch_1854_1, Williams_1970}
\begin{equation}\label{eq:kww}
    \varphi_{\text{\tiny{KWW}}}(t) = \exp[-\left(t/\tau\right)^{\beta}],
\end{equation}
and report the results in Table~\ref{tab:fragilities}. We plot the relaxation of $\varphi(t)$ in Fig.~\ref{fig:dj} as a function of $t$. The results of the fits to Eq.~\eqref{eq:kww} is reported in dashed lines. Different fragilities yield curves that would not superpose by simple time rescaling since they are described by KWW functions with different stretching exponents $\beta$. The results of these fits will be used in the linear response regime analysis described below.

\subsection{Kovacs hump}
We describe the memory dynamics by means of a normalized function $K(t)$ for the double-jump relaxation for $t>t_a$
\begin{equation}\label{eq:K}
  K(t) = \frac{E(t) - E_\infty}{E(0) - E_\infty},
\end{equation}
with a similar definition as $\varphi(t)$ for single-jump relaxation in Eq.~\eqref{eq:phi}. We report in Fig.~\ref{fig:humps}$(a)$, $(b)$, $(c)$ and $(d)$ the results for the four different values of $G_0$ and the five different annealing temperatures $T_a$ given by the ratios $\theta_a = T_a/T_f \in \{0.995, 0.990, 0.980, 0.960, 0.920\}$, against $(t - t_a)$ rescaled by $\langle\tau\rangle=4\tau/\pi$~\cite{Ruiz_Garc_a_2014}. As observed in experiments~\cite{Kovacs1964, Bernazzani_2002} the maximum height of the hump, $K_{\text{max}}$, grows as the annealing temperature $T_a$ lowers, and its time of occurrence $t_k$, shifts to smaller values closer to $t_a$. On the other hand, the effect of the fragility results in an earlier convergence~\cite{Aquino_2006} to the single down-jump curves reproduced here from Fig.~\ref{fig:dj}. In Fig.~\ref{fig:humps}$(e)$, $(f)$, $(g)$ and $(h)$ we report the same curves in log-log scales normalized by $K_{\text{max}}$ and with time rescaled by $t_k - t_a$, where $t_k$ is the time when the peak occurs: one can see a convergence to a similar hump shape independent of the annealing temperature~\cite{Prados2010_KovacsME} for each value of $G_0$.

Let us now briefly introduce the linear response analysis proposed in~\cite{Prados2010_KovacsME}: it has been shown that for systems whose dynamics can be modeled by a master equation, there exist an analytic prediction for the shape of the hump $K(t)$ as a function of the single down-jump relaxation $\varphi(t)$ in the linear response regime. In particular, the exact expression reads ~\cite{Prados2010_KovacsME, Ruiz_Garc_a_2014}
\begin{equation}\label{eq:linear_resp}
    K(t) = \frac{\varphi(t) - \varphi(t_a)\varphi(t - t_a)}{1 - \varphi(t_a)}.
\end{equation}
\begin{figure}[!t]
  \includegraphics[scale=0.3625]{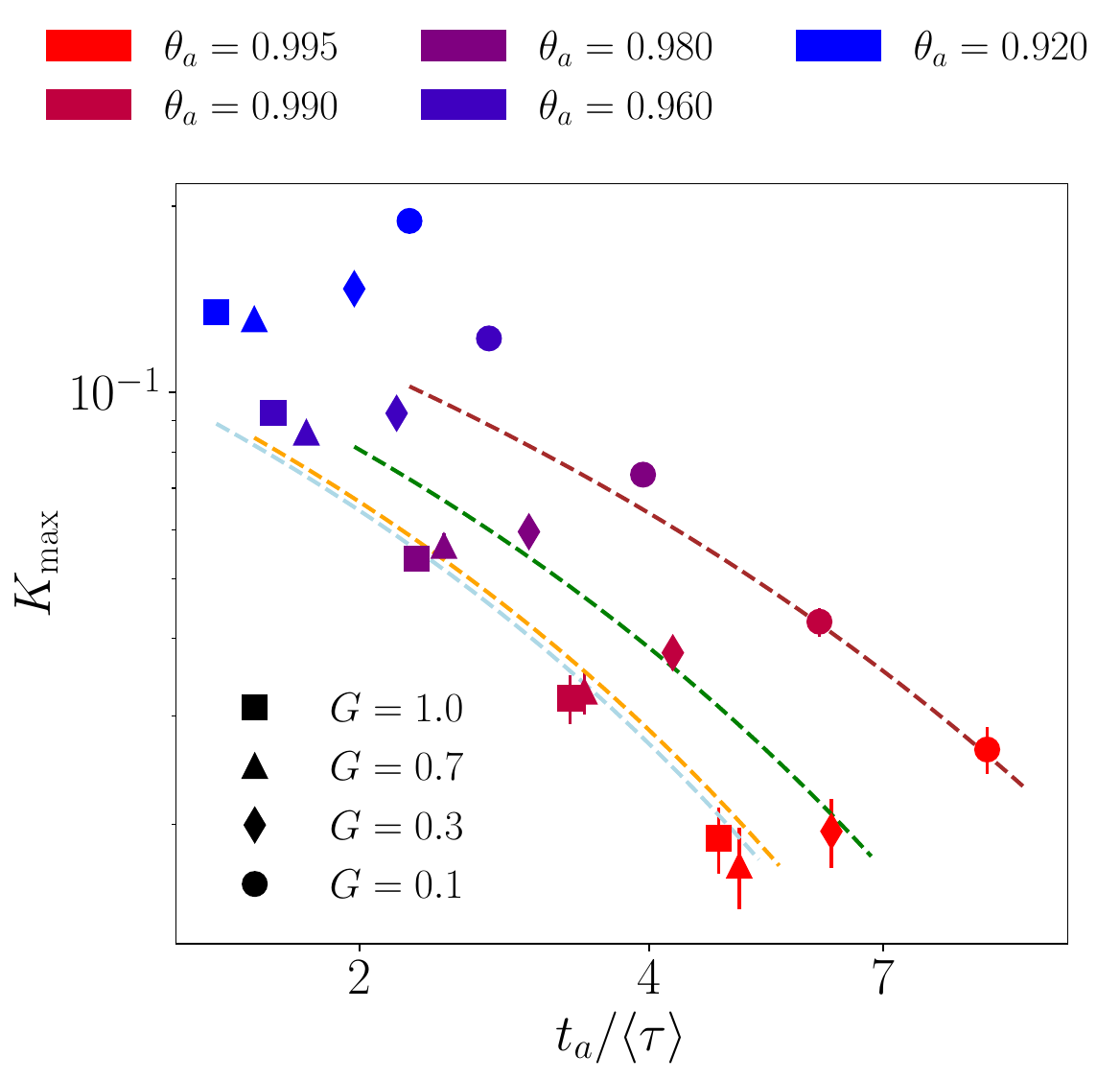}
  \caption{The hump peak height $K_{\text{max}}$ against normalized annealing time $t_a/\langle \tau \rangle$ in log-log scales for four values of $G_0$; dashed lines represent linear response result computed from Eq.~\eqref{eq:linear_resp}.}\label{fig:lin_resp}
\end{figure}As one can see from Fig.~\ref{fig:humps}$(a)$, $(b)$, $(c)$ and $(d)$ the solid lines computed from~\eqref{eq:linear_resp} using parameters from the KWW fits (see Table~\ref{tab:fragilities}), show a good match to the hump data (for $\theta_a \geq 0.960$). The agreement, however, worsens for lower values of the annealing temperature $T_a$ and for smaller values of $G_0$, i.e. for more fragile systems. In both cases the nonlinear effects are enhanced, thus explaining the departure from Eq.~\eqref{eq:linear_resp}. Furthermore, we studied the dependence of the hump height $K_{\text{max}}$ on the annealing time $t_a$~\cite{Ruiz_Garc_a_2014} by computing the maximum of~\eqref{eq:linear_resp} and comparing it to our simulations. The results are reported in Fig.~\ref{fig:lin_resp} displaying a good agreement with the linear response predictions (dashed lines) for higher annealing temperatures independently of the value of $G_0$. Interestingly, these results fall in a similar range of values as those obtained for the one-dimensional Ising model~\cite{Ruiz_Garc_a_2014}. Indeed, the DPLM data seem compatible with a power-law scaling between $K_{\text{max}}$ and $t_a$, but in the case of the present study, the range of values is not sufficiently large to establish the scaling reliably.

\subsection{Memory Encoded in Particle Interactions}
\begin{figure}[!t]
  \includegraphics[scale=0.41]{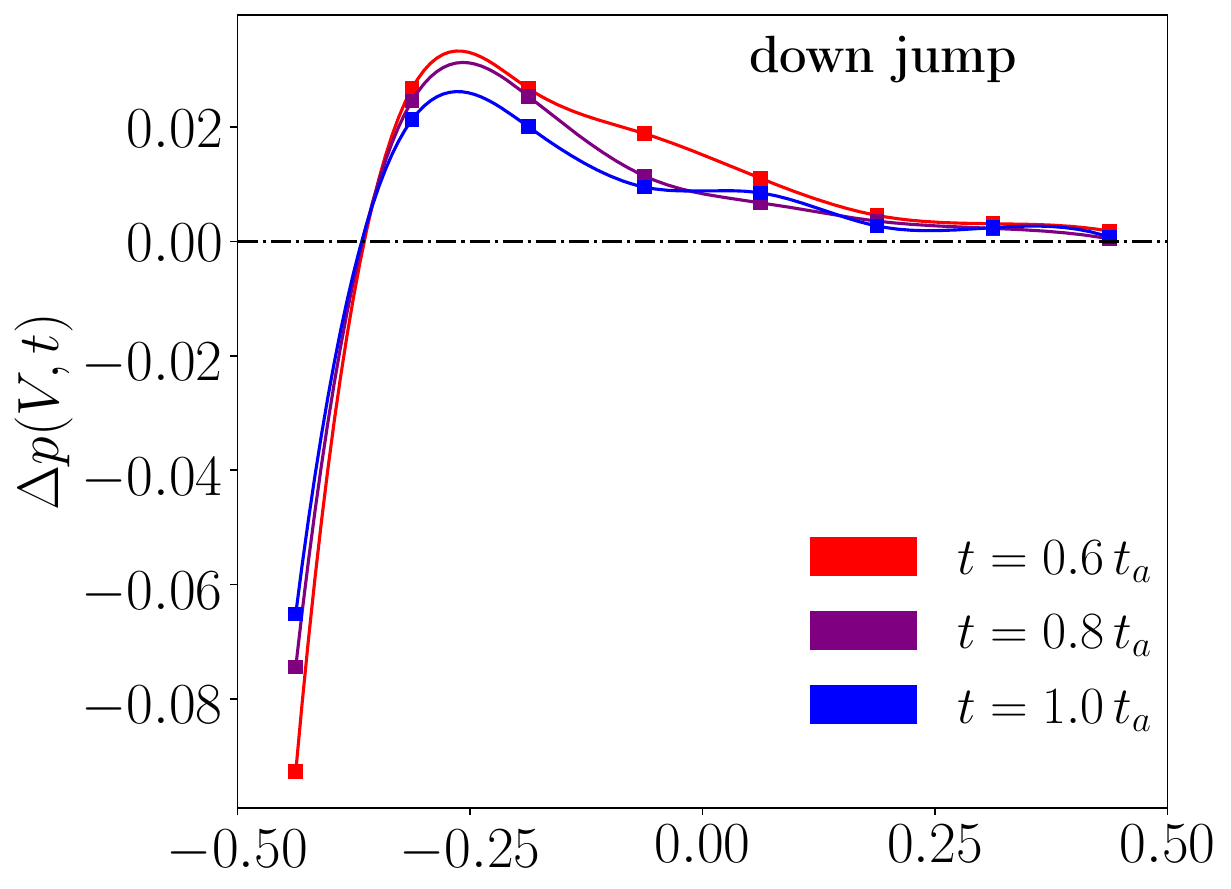}
  \caption{Sequence of $\Delta p(V, t)$ (see Eq.~\eqref{eq:DE}) for the down-jump dynamics at three different time $t$, all displaying only one isolated zero.
  }\label{fig:dp_histo_downj}
\end{figure}
\begin{figure}[!t]
  \includegraphics[scale=0.38]{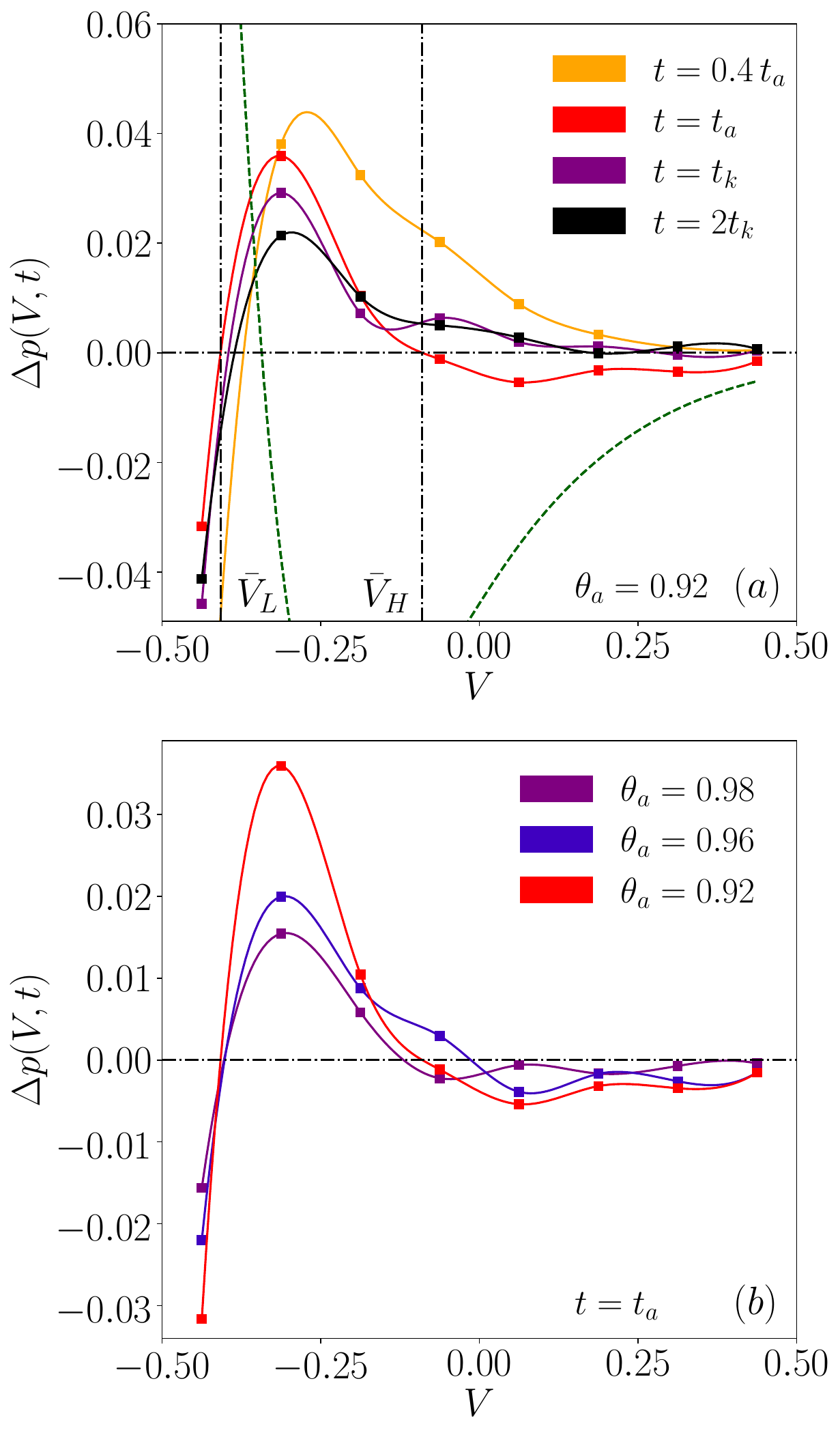}
  \caption{$(a)$: Sequence for the double-jump (memory) protocol for four different time $t$. For $t \ll t_a$, $\Delta p$ displays only one isolated zero, with a second one appearing at $t = t_a$. For $t \gg t_a$, the case of a single isolated zero is restored. The green dashed line represents $\peq(V, T_a) - \peq(V, T_f)$. $(b)$: $\Delta p$ at $t = t_a$ for different values of annealing temperature $T_a = \theta_a T_f$.
  }\label{fig:dp_histo_double}
\end{figure}
Let us look at the system from the perspective of the probability distribution $p(V,t)$ of the realized interactions $V_{s_i s_j}$ among particles at time $t$, which has revealed detailed information about the state of the system in previous studies \cite{DPLM2017,Lulli_2020}. By definition, at the initial and the final states, such distribution coincides with the equilibrium one, i.e., $p(V,0) = \peq(V,T_i)$ and $p(V,\infty) = \peq(V,T_f)$. We now examine its detailed
evolution. 
For the sake of the discussion we will consider $G_0 = 1$, but similar arguments can be used for $G_0 < 1$. Let us define 
\begin{equation}\label{eq:dp}
    \Delta p(V,t) = p(V,t) - \peq(V,T_f),
\end{equation}
which displays the difference of the distribution $p(V,t)$ with respect to the final equilibrium distribution at $T_f$.

We first study the simple case of a single temperature jump from $T_i$ to $T_f$, the relaxation of which has already been depicted in Fig.~\ref{fig:dj}. 
We report in Fig.~\ref{fig:dp_histo_downj} the evolution of $\Delta p(V,t)$. 
We observe that at all time $t$, $\Delta p(V,t)$ is positive for $V \agt -0.37$ and negative for $V \alt -0.37$, indicating that the distribution $p(V,t)$ is skewed towards high energy interactions compared with $\peq(V,T_f)$ due to the high initial temperature $T_i$.
Thus, the function $\Delta p(V,t)$ admits only one zero at $V\simeq -0.37$.
During cooling, $\Delta p(V,t)$ converges towards 0 for all $V$. To relate to the relaxation of the total energy, we consider the difference $\Delta E(t) = E(t) - E_\infty$ between the instantaneous system energy $E(t)$ and the equilibrium energy at $T_f$, analogous to definitions in Eqs. (\ref{eq:phi}) and (\ref{eq:K}).  It can be expressed as
\begin{equation}
  \label{eq:DE}
\Delta E(t) = 2 N \int \mbox{d} V ~ V \Delta p(V, t) 
\end{equation}
where $N$ is the number of particles in the systems and we have assumed a small void density $\phi_v\simeq 0$ for simplicity.
It is easy to see that a single isolated zero of $\Delta p(V,t)$ in Figure~\ref{fig:dp_histo_downj} indeed implies  ${\Delta E(t) > 0}$. 
To arrive at $\Delta E(t)=0$, a necessary condition of equilibrium, one then requires $\Delta p(V,t) \equiv 0$ for all $V$, implying an equilibrium distribution $\peq(V,T_f)$ of the interactions. Kovacs' condition of $E(t)=E_\infty$ therefore only occurs at equilibrium at long time for the single-jump case.

We now return to our main focus of the double temperature jump protocol. Figure~\ref{fig:dp_histo_double}$(a)$ shows $\Delta p(V,t)$ for $\theta_a = 0.92$. The initial evolution is similar to the single jump case.
However, at $t_a$, $p(V,t)$ at large $V$ has become close to $\peq(V,T_a)$ (see green dashed line which shows $\peq(V,T_a)-\peq(V,T_f)$ in Fig.~\ref{fig:dp_histo_double}$(a)$).
The better equilibration towards $T_f$ at large $V$ is due to the generally faster dynamics of the weakly bonded particles. As a result, $p(V,t)<0$ at large $V$, leading to two zeros of $p(V,t)$ at $\bar{V}_L\simeq -0.41$ and $\bar{V}_H\simeq -0.09$. Interestingly, the addition of a zero allows the system to satisfy Kovacs' condition of $\Delta E(t_a)=0$ even when $\Delta p(V,t_a)\not\equiv 0$, as is explicitly illustrated in Fig.~\ref{fig:dp_histo_double}$(a)$. We also report in Fig.~\ref{fig:dp_histo_double}$(b)$ the data for higher values of $T_a$ at $\theta_a = 0.92, 0.96$ and $0.98$ and are all consistent with having two isolated zeros.
The non-vanishing $\Delta p(V,t_a)$ despite $\Delta E(t_a)=0$ evidences a non-equilibrium interaction distribution $p(V, t_a)$, which is the microscopic origin of the material memory responsible for the Kovacs hump.

In Fig.~\ref{fig:dp_histo_double}$(a)$, we have also plotted $\Delta p(V,t)$  at $t > t_a$ after the second temperature jump to $T_f$. For large $V$, $\Delta p(V,t)$ rises to 0 rapidly. It corresponds to warming of these overly-cooled interactions. This is the dominant mechanism of the return of $\Delta E(t)$ to a positive value at the Kovacs hump. It also restores the usual case of a single isolated zero of $\Delta p(V,t)$ so that subsequent equilibration proceeds in a way similar to the single-jump.

\subsection{Mobile and immobile particles}
\begin{figure}[!t]
  \includegraphics[scale=0.41]{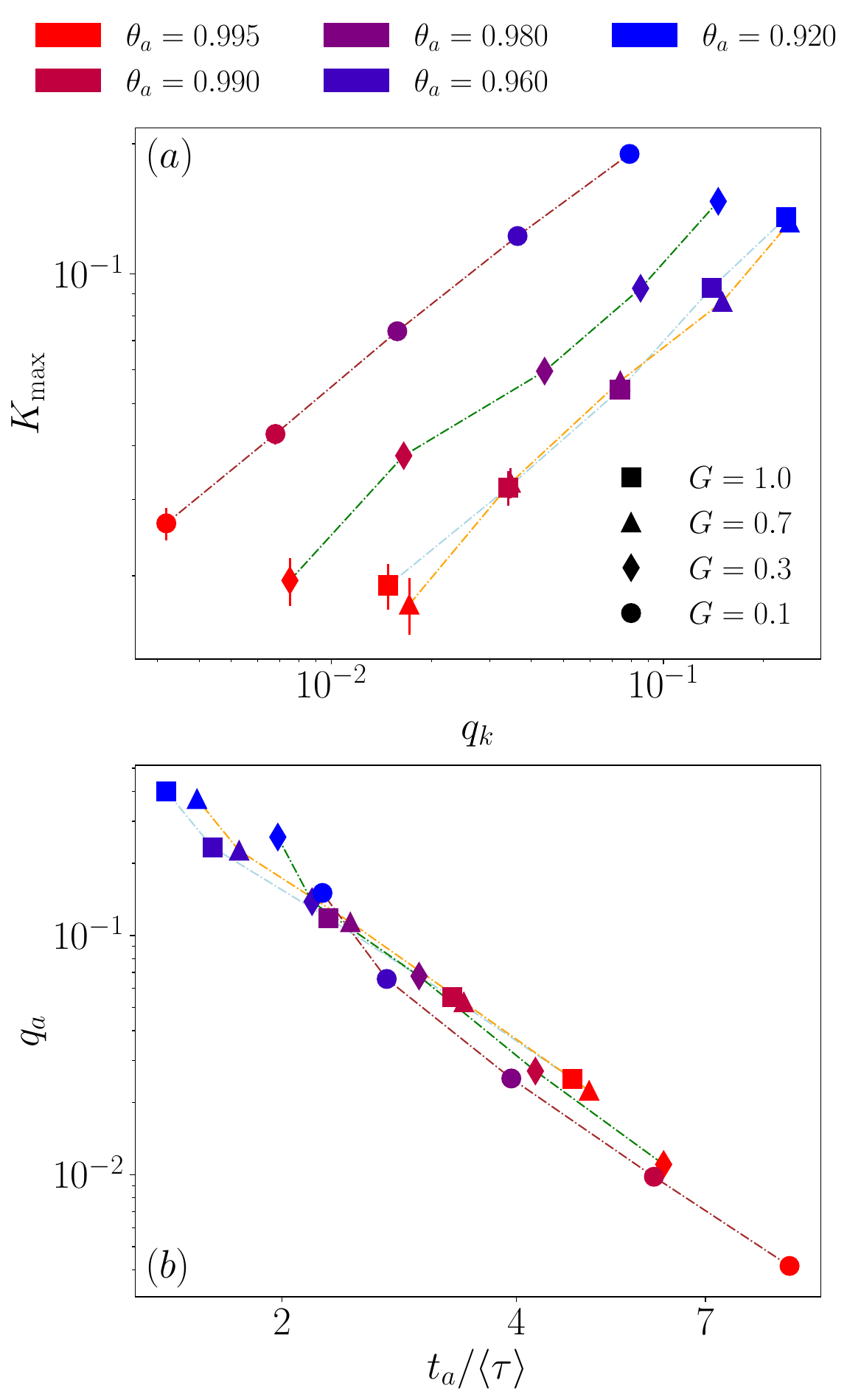}
  \caption{$(a)$: $K_{\text{max}}$ against the fraction of immobile particles $q_k = q(t_k)$ at the peak of the hump. $(b)$: fraction of immobile particles $q_a = q(t_a)$ against normalized annealing time $t_a/\langle \tau \rangle$; data superpose well for different fragilities, i.e. different values of $G_0$.}\label{fig:k_and_q}
\end{figure}
\begin{figure}[!t]
  \includegraphics[scale=0.38]{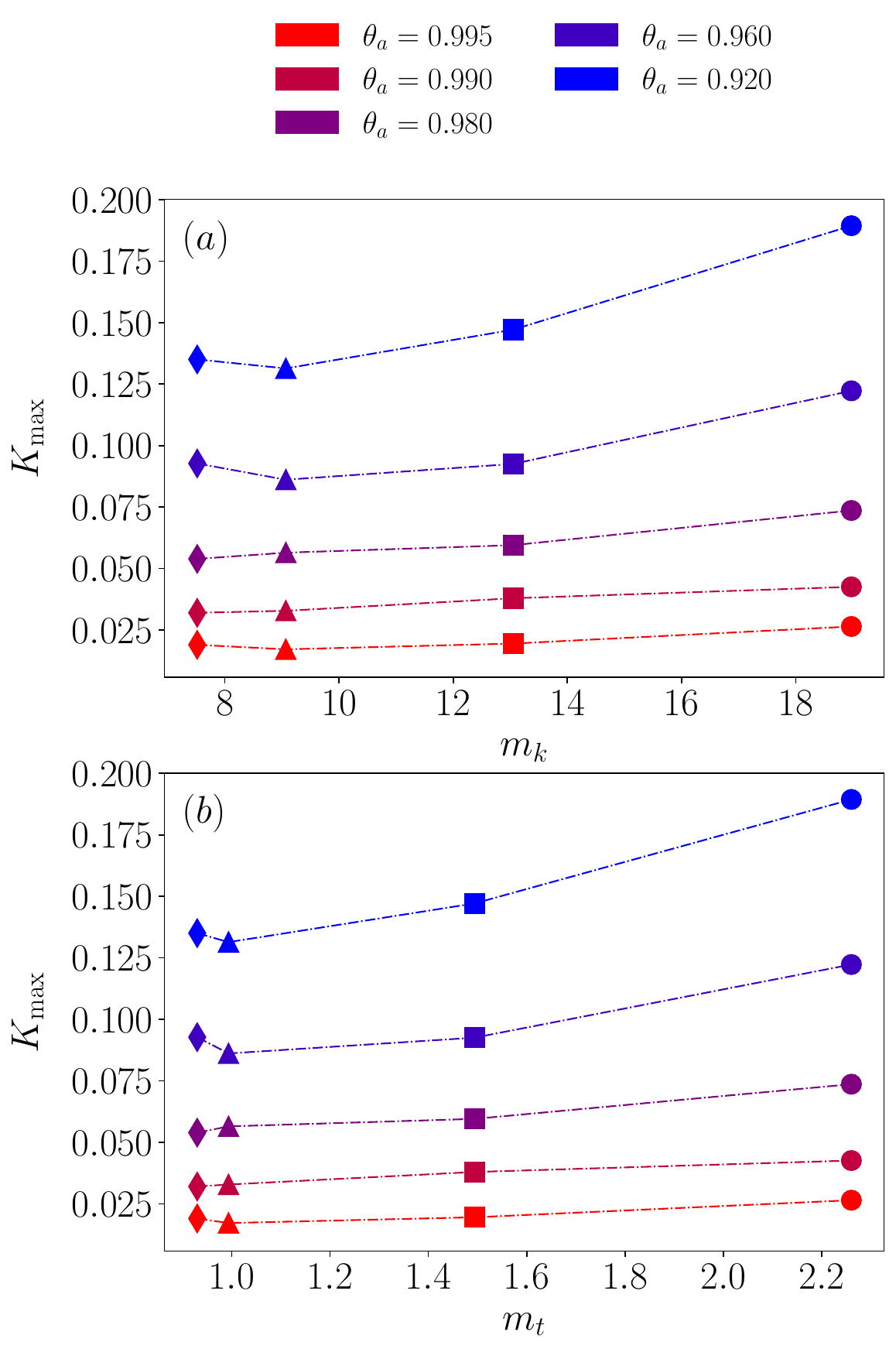}
  \caption{Peak values of the Kovacs hump $K_{\text{max}}$ against the kinetic fragility $m_k$ in $(a)$, and thermodynamic fragility $m_t$ in $(b)$. From left to right the corresponding values of $G_0$ are $0.1, 0.3, 0.7$ and $1.0$. Dashed lines connect points sharing the same normalized annealing temperature $\theta_a=T_a/T_f$.}\label{fig:hump_vs_m}
\end{figure}
The DPLM allows us to study of the dynamics by grouping particles according to their mobility. One can measure the evolution of the system by introducing an \emph{overlap} parameter $q(t)$~\cite{Lulli_2020} which measures the fraction of particles that are located at their initial position at time $t$, hence $0 \leq q(t) \leq 1$ with $q(t) = 0$ when all particles are away from their initial positions. 

We plot in Fig.~\ref{fig:k_and_q}$(a)$ the maximum of the hump $K_{\text{max}} = K(t_k)$ against the corresponding value of the overlap $q_k = q(t_k)$. As seen, $K_{\text{max}}$ scales with $q_k$ and the strongest glass yields the largest value of the overlap $q_k$. The hump is taller with a larger $q_k$, for lower annealing temperatures. 

Another possible comparison between the different systems can be obtained by looking at the overlap at the annealing time $q_a = q(t_a)$ as a function of $t_a$. At this time, for each value of $G_0$, all systems are characterized by the same average energy $E_\infty$. We report these results in Fig.~\ref{fig:k_and_q}$(b)$: the data collapse on the same curve indicating that, for the prescribed rescaled annealing time $t_a/\langle\tau\rangle$, the ratio of mobile/immobile particles in the system does not depend on the degree of fragility of the system. It is possible to notice that, for a fixed value of $G_0$, a lower annealing temperature corresponds to a larger $q_a$ and a smaller $t_a/\langle \tau\rangle$, signaling that the faster dynamics is driven by an increasingly smaller fraction of mobile particles.

We conclude our analysis of the Kovacs hump by reporting in Fig.~\ref{fig:hump_vs_m} the value of $K_{\text{max}}$ as a function of both the kinetic and thermodynamic fragility indices $m_k$ and $m_t$~\cite{Lee_2020}, clearly showing a taller hump as the fragility increases. To the best of our knowledge such an analysis is lacking in the experimental literature. However, it would be very interesting to examine if such a trend may exist in experimental data.

\begin{figure}[!t]
  \includegraphics[scale=0.3]{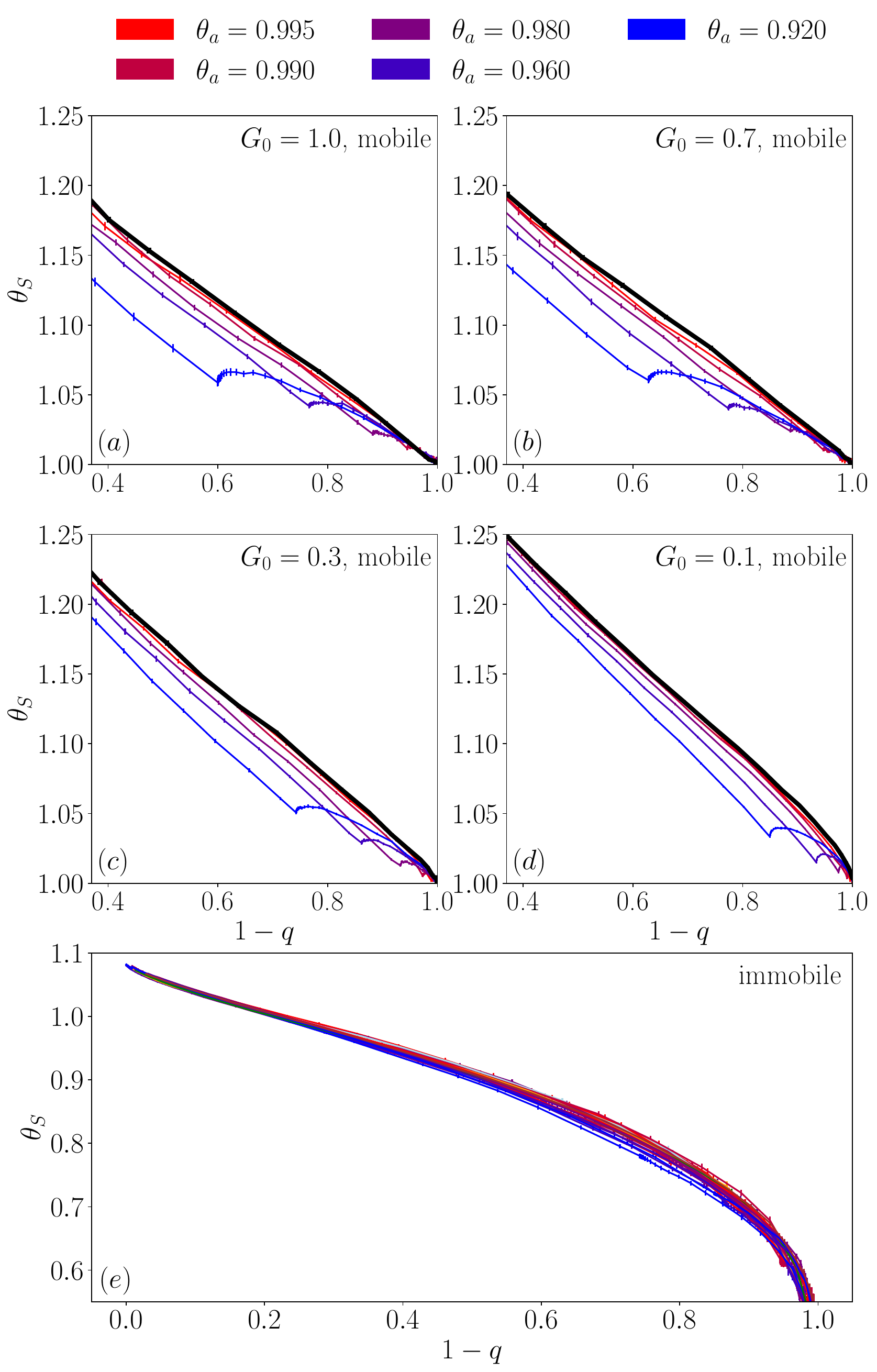}
  \caption{Panels $(a)$ to $(d)$: Normalized structural temperature $\theta_S = T_S/T_f$ for mobile regions as a function of the fraction of mobile particles $1-q$; single down-jump relaxation data are reported in black solid lines. Panel $(e)$: Normalized structural temperature for the immobile regions for different $G_0$ and annealing temperature $T_a$.}\label{fig:st}
\end{figure}

\subsection{Structural Temperature}
Finally, we analyze the \emph{structural temperature}~\cite{Lulli_2020} $T_S$, which, analogous to the fictive temperature, describes the effective temperature of the particle interactions. It is computed by solving numerically $E(t) = \int_{V_0}^{V_1}\mbox{d}V\,V\,\peq(V, T_S(t))$. In~\cite{Lulli_2020} it was shown that the local value of $T_S$ strongly correlates with the mobility of the particles. In order to compare systems at different fragilities we use the ratio $\theta_S = T_S/T_f$ and we study its temporal evolution for both mobile and immobile regions, using $1 - q$ as an evolution parameter that grows as time passes, eventually reaching unity. Results are reported in Fig.~\ref{fig:st}: panels $(a)$, $(b)$, $(c)$ and $(d)$ all display the evolution of the structural temperature $T_S$ in the mobile regions with a cusp at the value $1 - q_a$, corresponding to the annealing time $t_a$; panel $(e)$ displays $\theta_S$ for the immobile regions where, however, it is not possible to notice any significant feature at $1 - q_a$. The evolution of $T_s$ at the mobile and immobile regions is due to both particle motions in the mobile regions and the growth of the mobile regions at the expense of the immobile ones by shrinking their boundaries. 

The cusps exhibited by the mobile particles indicates the reheating of the overly cooled and weakly bonded particles as explained above. In addition, one can see a remarkable convergence of the dynamics of the immobile regions independently on the value of $G_0$. Values of $\theta_a$ as low as 0.6 can be reached because, at the end of the evolution of the immobile regions, mostly the lowest interaction energies close to $V_0$ will be left. At the end of its evolution the immobile regions are mostly composed of highly stable configurations, yielding a very low average energy, hence a very low structural temperature.

\section{Conclusions}\label{sec:conclusions}
In this paper we demonstrated the ability of the Distinguishable Particles Lattice Model (DPLM) to capture the main features of the Kovacs ageing dynamics. In particular, we studied systems with different~\emph{a priori} distributions of the interaction energies (see Eq.~\eqref{eq:g}) which are related to different fragilities~\cite{Lee_2020}: we analyzed the Kovacs memory dynamics in a broader range of fragilities than what was previously done in the literature~\cite{Aquino_2006}. In this extended setting we studied in detail the Kovacs memory response in the linear regime obtaining a good agreement with the master equation approach detailed in~\cite{Prados2010_KovacsME} for annealing temperature $T_a$ satisfying $1 - T_a/T_f = 1 - \theta_a \leq 2\times 10^{-2}$.

We identified the memory content of the system in the features of the function $\Delta p(V,t) = p(V,t) - \peq(V, T_f)$, i.e. the difference between the instantaneous distribution and the final equilibrium one of the particle     interaction energies. In particular, after the annealing, $\Delta p(V,t_a)$ yields two zeros allowing the system to have the same equilibrium energy as at $T_f$, i.e. $E(t_a) = E_\infty$, while displaying an out-of-equilibrium distribution. The subsequent heating of the system at $T_f$ repopulates the overly cooled high-energy part of the distribution yielding the Kovacs hump and restoring the scenario of a single isolated zero. At this point the relaxation continues as in the simple down-jump case.

Further, we characterized the hump height $K_{\max}$ from a particle-mobility perspective. A clear correlation between $K_{\max}$ and the fragility index, both kinetic and thermodynamic, was provided. Also, the fraction of immobile particles at the end of the annealing $q_a = q(t_a)$ was found to collapse to the same function of $t_a/\langle \tau \rangle$, independent of the fragility of the system. Finally, we studied the dynamics of the structural temperature~\cite{Lulli_2020}, for both mobile and immobile regions, as a function of the fraction of mobile particles $1-q$: this illustrates that the Kovacs' hump is associated with the evolution of $T_S$ in the mobile regions while the dynamics in the immobile ones is weakly dependent on the fragility of the system.

\section{Acknowledgements}
We gratefully acknowledge Haihui Ruan and Giorgio Parisi for interesting discussion and comments. We thank the support of the National Natural Science Foundation of China (Grant No.~11974297,~12050410244).

\bibliographystyle{apsrev4-1}
\bibliography{references}

\end{document}